\documentstyle[12pt]{article}
\begin{document}
\begin{titlepage}

\title{The angular momentum of the gravitational field
and the Poincar\'{e} group}

\author{J. W. Maluf$\,^{*}$, S. C. Ulhoa, F. F. Faria\\
Instituto de F\'{\i}sica, \\
Universidade de Bras\'{\i}lia\\
C. P. 04385 \\
70.919-970 Bras\'{\i}lia DF, Brazil\\
and\\
J. F. da Rocha-Neto$\,^{\S}$\\
Universidade Federal do Tocantins\\
Campus Universit\'ario de Arraias\\
77.330-000 Arraias TO, Brazil\\}
\date{}
\maketitle

\begin{abstract}
We redefine the gravitational angular momentum in the framework of the
teleparallel equivalent of general relativity. In similarity to the
gravitational energy-momentum, the new definition for the gravitational
angular momentum is coordinate independent. By considering the 
Poisson brackets in the phase space of the theory, we find that the
gravitational energy-momentum and angular momentum correspond to a
representation of the Poincar\'e group. This result allows us to 
define Casimir type invariants for the gravitational field.
\end{abstract}
\thispagestyle{empty}
\vfill
\noindent PACS numbers: 04.20.Cv, 04.20.Fy\par
\bigskip
\noindent (*) e-mail: wadih@fis.unb.br\par
\noindent ($\S$) e-mail: rocha@uft.edu.br
\end{titlepage}
\newpage

\noindent

\section{Introduction}

The teleparallel equivalent of general relativity (TEGR) is a viable
alternative geometrical description of Einstein's general relativity
in terms of the tetrad field \cite{Hehl}, and continues to be 
object of thorough investigations \cite{Obukhov,Itin1,Itin2,OR}. 
In the framework of the
TEGR it has been possible to address the longstanding problem of 
defining the energy, momentum and angular momentum of the 
gravitational field \cite{Maluf1,Maluf2,Maluf3}. The tetrad field 
seems to be a suitable field quantity to address this problem,
because it yields the gravitational field and at the same time 
establishes a class of reference frames in space-time \cite{Maluf4}.
Moreover there are simple
and clear indications that the gravitational energy-momentum
defined in the context of the TEGR provides a unified picture of the
concept of mass-energy in special and general relativity. 

In special
relativity the energy of an arbitrary body is frame dependent, and a
similar property is expected to hold in general relativity. For
instance, a black hole of mass $m$ that is distant from an observer 
behaves as a particle of mass $m$. If the observer is at rest 
with respect to the black hole, he or she will conclude that
the energy of the black hole is $mc^2$. If, however, the observer is
moving at a velocity $v$ with respect to the black hole, then for 
this observer the black hole energy is $\gamma mc^2$ (asymptotically),
where $\gamma=(1-v^2/c^2)^{-1/2}$. The latter expression 
establishes the frame dependence of the black hole energy. This 
feature is naturally exhibited by the global SO(3,1) covariance of 
the gravitational energy-momentum defined in the framework of the 
TEGR \cite{Maluf5}. The frame dependence of the gravitational energy
is not restricted to its total energy, neither to asymptotic regions.
It holds in the consideration of the gravitational energy over 
finite regions of the three-dimensional spacelike hypersurface of
an arbitrary space-time (de Sitter or anti-de Sitter space-times, for 
instance). As an example of a finite region mentioned above, we 
may consider the gravitational energy contained within the (external) 
event horizon of a black hole, a quantity that defines its irreducuble
mass. We note that the dependence of the total Schwarzschild mass with
respect to a boosted reference system has been addressed previously by 
York \cite{York}, who constructed the boosted Schwarzschild initial 
data by means of a suitable coordinate transformation in the 
Schwarzschild space-time.

In this article we redefine the angular momentum of the
gravitational field in the framework of the TEGR. In similarity
to the definition of the gravitational energy-momentum, we 
interpret the appropriate constraint equations as equations that 
define the gravitational angular momentum. The latter turns out to be
coordinate independent. We will show that the energy-momentum and
angular momentum of the gravitational field satisfy the Poincar\'e
algebra in the phase space of the theory. As a consequence of this
result, we may define Casimir type invariants for an arbitrary 
configuration of the gravitational field. 

The present definition is conceptually different from previous
approaches to the gravitational angular momentum. The definition
arises by considering the field equations of the theory. In
contrast, other approaches are based on boundary terms to the
Hamiltonian or to the action integral. Regge and Teitelboim
\cite{RG} obtained a Hamiltonian formalism for general relativity
that is manifestly invariant under Poincar\'{e} transformations at
infinity by introducing ten new pairs of canonical variables, which
yield ten surface integrals to the total Hamiltonian. The subsequent
analysis by York \cite{York} showed that a proper definition for the
gravitational angular momentum requires a suitable asymptotic 
behaviour of the spatial components of the Ricci tensor (that 
ensures an ``almost" rotational symmetry of the spatial components
$g_{ij}$ of the metric tensor).  
A careful analysis of the exact form of the boundary conditions 
needed to define the energy, momentum and angular momentum of the
gravitational field has been carried out by Beig and \'{o}
Murchadha \cite{BM}, and by Szabados \cite{Sza}, who found the
necessary conditions that yield a finite value for the above
mentioned quantities. In these analises the Poincar\'{e} 
transformations and the Poincar\'{e} algebra are
realized at the spacelike infinity. These are transformations
of the Cartesian coordinates in the asymptotic region of the 
space-time. 

In the present analysis we show that the Poincar\'{e} algebra is
realized in the full phase space of the TEGR. We  
evaluate the angular momentum of a simple configuration of the 
gravitational field, namely, of the space-time of a rotating 
mass shell, in order to display the procedure regarding the 
present definition. We will pay special attention to the frame
dependence of the latter.

\bigskip
Notation: space-time indices $\mu, \nu, ...$ and SO(3,1)
indices $a, b, ...$ run from 0 to 3. Time and space indices are
indicated according to
$\mu=0,i,\;\;a=(0),(i)$. The tetrad field is denoted by $e^a\,_\mu$,
and the torsion tensor reads
$T_{a\mu\nu}=\partial_\mu e_{a\nu}-\partial_\nu e_{a\mu}$.
The flat, Minkowski space-time metric tensor raises and lowers
tetrad indices and is fixed by
$\eta_{ab}=e_{a\mu} e_{b\nu}g^{\mu\nu}= (-+++)$. The determinant of 
the tetrad field is represented by $e=\det(e^a\,_\mu)$.\par        
\bigskip

\section{The Lagrangian and Hamiltonian formulations of the TEGR}

We will briefly recall both the Lagrangian and Hamiltonian 
formulations of the TEGR. The Lagrangian density for the 
gravitational field in the TEGR is given by

\begin{eqnarray}
L(e_{a\mu})&=& -k\,e\,({1\over 4}T^{abc}T_{abc}+
{1\over 2} T^{abc}T_{bac} -T^aT_a) -L_M\nonumber \\
&\equiv&-k\,e \Sigma^{abc}T_{abc} -L_M\;,
\label{1}
\end{eqnarray}
where $k=1/(16 \pi)$, and $L_M$ stands for the Lagrangian density
for the matter fields. As usual, tetrad fields convert space-time 
into Lorentz indices and vice-versa. 
The tensor $\Sigma^{abc}$ is defined by

\begin{equation}
\Sigma^{abc}={1\over 4} (T^{abc}+T^{bac}-T^{cab})
+{1\over 2}( \eta^{ac}T^b-\eta^{ab}T^c)\;,
\label{2}
\end{equation}
and $T^a=T^b\,_b\,^a$. The quadratic combination
$\Sigma^{abc}T_{abc}$ is proportional to the scalar curvature
$R(e)$, except for a total divergence. The field
equations for the tetrad field read

\begin{equation}
e_{a\lambda}e_{b\mu}\partial_\nu(e\Sigma^{b\lambda \nu})-
e(\Sigma^{b \nu}\,_aT_{b\nu \mu}-
{1\over 4}e_{a\mu}T_{bcd}\Sigma^{bcd})
\;= {1\over {4k}}eT_{a\mu}\,.
\label{3}
\end{equation}
where $eT_{a\mu}=\delta L_M / \delta e^{a\mu}$. 
It is possible to prove by explicit calculations that the left hand
side of Eq. (3) is exactly given by ${1\over 2}\,e\,
\lbrack R_{a\mu}(e)-{1\over 2}e_{a\mu}R(e)\rbrack$.
The field equations above may be rewritten in the form 

\begin{equation}
\partial_\nu(e\Sigma^{a\lambda\nu})={1\over {4k}}
e\, e^a\,_\mu( t^{\lambda \mu} + T^{\lambda \mu})\;,
\label{4}
\end{equation}
where

\begin{equation}
t^{\lambda \mu}=k(4\Sigma^{bc\lambda}T_{bc}\,^\mu-
g^{\lambda \mu}\Sigma^{bcd}T_{bcd})\,,
\label{5}
\end{equation}
is interpreted as the gravitational energy-momentum tensor
\cite{Maluf3}. (We remark that an energy-momentum tensor for
cosmological perturbations has been considered in the framework
of the Hilbert-Einstein formulation \cite{PetrovKatz}. Such
energy-momentum tensor is defined with respect to certain
backgrounds, and is related to Einstein's equations via
differential conservation laws.)

The Hamiltonian formulation of the TEGR is obtained by first 
establishing the phase space variables. The Lagrangian density
does not contain the time derivative of the tetrad component 
$e_{a0}$. Therefore this quantity will arise as a Lagrange 
multiplier. The momentum canonically conjugated to $e_{ai}$ is 
given by $\Pi^{ai}=\delta L / \delta \dot{e}_{ai}$. The 
Hamiltonian formulation is obtained by rewriting the Lagrangian
density in the form $L=p\dot{q}-H$, in terms of $e_{ai}$, 
$\Pi^{ai}$ and Lagrange multipliers. The Legendre transform
can be successfuly carried out, and the final form of the 
Hamiltonian density reads  \cite{Maluf6}

\begin{equation}
H=e_{a0} C^a+ \alpha_{ik}\Gamma^{ik} + \beta_k \Gamma^k\,,
\label{6}
\end{equation}
plus a surface term. $\alpha_{ik}$ and $\beta_k$ are Lagrange
multipliers that (after solving the field equations) are identified
as $\alpha_{ik}=1/2(T_{i0k}+T_{k0i})$ and $\beta_k=T_{00k}$.
$C^a$, $\Gamma^{ik}$ and $\Gamma^k$ are first class constraints.
The Poisson brackets between any two field quantities $F$ and $G$
is given by

\begin{equation}
\lbrace F,G\rbrace=\int d^3x \biggl(
{{\delta F}\over {\delta e_{ai}(x)}}
{{\delta G}\over {\delta\Pi^{ai}(x)}}-
{{\delta F}\over {\delta\Pi^{ai}(x)}}
{{\delta G}\over {\delta e_{ai}(x)}} \biggr)\;,
\label{7}
\end{equation}
We recall that the Poisson brackets 
$\lbrace \Gamma^{ij}(x),\Gamma^{kl}(y)\rbrace$
reproduce the angular momentum algebra \cite{Maluf6}.

The constraint $C^a$ is written as 
$C^a=-\partial_i \Pi^{ai}+ h^a$, where $h^a$ is an intricate 
expression of the field variables. The integral form of the
constraint equation $C^a=0$ motivates the definition of the
gravitational energy-momentum tensor $P^a$ \cite{Maluf1},

\begin{equation}
P^a=-\int_V d^3x \partial_i \Pi^{ai}\,.
\label{8}
\end{equation}
$V$ is an arbitrary volume of the three-dimensional space. In the
configuration space we have

\begin{equation}
\Pi^{ai}=-4ke \Sigma^{a0i}\,.
\label{9}
\end{equation}
The emergence of total divergences in the form of scalar or vector 
densities is possible in the framework of theories constructed out 
of the torsion tensor. Metric theories of gravity do not share this 
feature.
We note that by making $\lambda=0$ in eq. (4) and identifying 
$\Pi^{ai}$ in the left hand side of the latter, the integral form of
eq. (4) is written as

\begin{equation}
P^a = \int_V d^3x \,e\,e^a\,_\mu(t^{0\mu}+ T^{0\mu})\,.
\label{10}
\end{equation}

It is important to rewrite the Hamiltonian density $H$ in the most 
simple form. We believe that the constraint $C^a$ admits a 
simplification, although we have not achieved it yet. However, we 
have been able to simplify the constraints 
$\Gamma^{ik}$ and $\Gamma^k$, which may be rewritten as a
single constraint $\Gamma^{ab}$. It is not difficult to verify that
the Hamiltonian density (6) may be written in the equivalent form

\begin{equation}
H=e_{a0}C^a+{1\over 2} \lambda_{ab} \Gamma^{ab}\,,
\label{11}
\end{equation}
where
$\lambda_{ab}=-\lambda_{ba}$ are Lagrange multipliers that are 
identified as $\lambda_{ik}=\alpha_{ik}$ and 
$\lambda_{0k}=-\lambda_{k0}=\beta_k$.  
$\Gamma^{ab}=-\Gamma^{ba}$ embodies both constraints 
$\Gamma^{ik}$ and $\Gamma^k$ by means of the relations
$\Gamma^{ik}=e_a\,^i e_b\,^k\Gamma^{ab}$, 
$\Gamma^k\equiv \Gamma^{0k} =e_a\,^0 e_b\,^k\Gamma^{ab}$. It reads

\begin{equation}
\Gamma^{ab}=M^{ab}+4ke(\Sigma^{a0b}-\Sigma^{b0a})\,,
\label{12}
\end{equation}
with $M^{ab}=e^a\,_\mu e^b\,_\nu M^{\mu\nu}=-M^{ba}$;
$M^{\mu\nu}$ is defined by

\begin{eqnarray}
M^{ik}&=&2\Pi^{\lbrack ik \rbrack}=e_a\,^i \Pi^{ak}-
e_a\,^k \Pi^{ai}\,, \\
M^{0k}&=&\Pi^{0k}=e_a\,^0 \Pi^{ak}\,. 
\label{13,14}
\end{eqnarray}

In similarity to the definition of $P^a$, the integral form of 
the constraint equation $\Gamma^{ab}=0$ motivates the new definition 
of the space-time angular momentum (in the expression previously
defined \cite{Maluf1}, $2\Pi^{\lbrack ik \rbrack}$ was taken as the
gravitational angular momentum density). The equation $\Gamma^{ab}=0$ 
implies 

\begin{equation}
M^{ab}=-4ke(\Sigma^{a0b}-\Sigma^{b0a})\,. 
\label{15}
\end{equation}
Therefore we define

\begin{equation}
L^{ab}=\int_V d^3x\; e^a\,_\mu e^b\,_\nu M^{\mu\nu}\,,
\label{16}
\end{equation}
as the 4-angular momentum of the gravitational field. In contrast to 
the definition presented in ref. \cite{Maluf1}, the expression above
is invariant under coordinate transformations of the three-dimensional
space. We note that on the right hand side of Eq. (15), as well as on
the right hand side of Eq. (9), there arises the time index $0$. We 
further note the presence of the determinant $e$ of the tetrad field 
in these quantities. This determinant can always be written as the 
product of the lapse function $N=(g^{00})^{-1/2}$ with the determinant 
of the triads restricted to the three-dimensional space 
$\,^3e$. Thus, $e=N(\,^3e)$. Because of 
the presence of the lapse function and of the time index $0$, the
right hand sides of Eqs. (9) and (15) are invariant under time
reparametrizations. 

Therefore $P^a$ and $L^{ab}$ are separately 
invariant under general coordinate transformations of the 
three-dimensional space and under time reparametrizations, which is 
an expected feature since these definitions arise in the Hamiltonian
formulation of the theory. Moreover these quantities transform 
covariantly under global SO(3,1) transformations.

We emphasize that expressions (8) and (16) are defined in the phase
space of the theory. In order to evaluate these expressions for a
particular field configuration, we consider the right hand side of 
Eqs. (9) and (15) in the configuration space of the theory.

\section{The Poincar\'{e} structure in the phase space of the theory}

An interesting result that follows from the definition above for the 
gravitational 4-angular momentum is that $L^{ab}$ and $P^a$ satisfy the
algebra of the Poincar\'e group. By means of the Poisson bracket 
defined by eq. (7) we find that the definitions (8), (13) (14) and 
(16) yield 

\begin{eqnarray}
\lbrace P^a , P^b \rbrace &=& 0\,, \nonumber \\
\lbrace P^a , L^{bc} \rbrace &=& -\eta^{ab} P^c+\eta^{ac} P^b 
\,, \nonumber \\
\lbrace L^{ab}, L^{cd} \rbrace &=&
-\eta^{ac}L^{bd} -\eta^{bd}L^{ac} +\eta^{ad}L^{bc}+\eta^{bc}L^{ad}
\,.
\label{17}
\end{eqnarray}
Equations (17) may be easily verified by considering the following
functional derivatives,

\begin{eqnarray}
{{\delta L^{ab}}\over {\delta e_{ck}(z)}}&=&
\int d^3 x {\delta \over {\delta e_{ck}(z)}}\biggl[
e^a\,_\mu e^b\,_\nu M^{\mu\nu}\biggr] \nonumber \\
&=& \int d^3 x {\delta \over {\delta e_{ck}(z)}}\biggl[
e^a\,_0 e^b\,_j\, M^{0j}+e^a\,_j e^b\,_0\, M^{j0}+
e^a\,_i e^b\,_j\, M^{ij}\biggr] \nonumber \\
&=&( \eta^{bc}e^a\,_0(z) -\eta^{ac} e^b\,_0(z))M^{0k}(z)
\nonumber \\
&{}&+(\eta^{bc} e^a\,_j(z)-\eta^{ac} e^b\,_j(z))\Pi^{kj}(z)
\nonumber \\
&{}&+(\eta^{ac} e^b\,_j(z)-\eta^{bc} e^a\,_j(z))M^{kj}(z)
\nonumber \\
&=&-\eta^{ac} \Pi^{bk}(z) + \eta^{bc} \Pi^{ak}(z)\,, 
\nonumber \\
{{\delta L^{ab}}\over {\delta \Pi_{ck}(z)}}&=&
\delta^a_c e^b\,_k(z)- \delta^b_c e^a\,_k(z)\,,
\nonumber \\
{{\delta P^a} \over {\delta e_{ck}(z)}}&=&0\,, \nonumber \\
{{\delta P^a} \over {\delta \Pi_{ck}(z)}}&=&-
\int d^3x \delta^a_c {\partial \over {\partial x^k}}
\delta^3(x-z)\,.
\label{18}
\end{eqnarray}

We see that the gravitational energy-momentum and angular
momentum constitute a representation of the Poincar\'e group.
It is well known that the field quantities that satisfy the algebra 
above are intimately related to energy-momentum and angular momentum.
Thus the Poincar\'e algebra of $P^a$ and $L^{ab}$ confirms the 
consistency of the definitions. 

\section{Tetrad fields as reference frames in space-time and the
gravitational angular momentum}

The theory defined by Eq. (1) is invariant under general coordinate
and global SO(3,1) transformations. Because of the global SO(3,1)
invariance of the theory, two tetrad fields that (i) are solutions of 
the field equations, (ii) yield the same metric tensor and (iii) are 
not related by a global SO(3,1) transformation, describe the
same space-time from the point of view of inequivalent reference 
frames. In view of this fact we must take into account the physical
and geometrical meaning of tetrad fields as reference frames adapted 
to ideal observers in space-time.

Before we proceed, we recall that the quadratic combination of the 
torsion tensor in Eq. (1) is equivalent to the scalar curvature 
except for a total divergence. The Lagrangian density (1) is 
invariant under infinitesimal Lorentz transformations only in the
context of asymptotically flat space-times, in which case the total 
divergence plays no role, and provided both the tetrad field and the
Lorentz transformation matrix satisfy appropriate boundary
conditions \cite{Cho}. In the general case, the Lagrangian density 
(1) is not invariant under local SO(3,1) transformations.

Each set of tetrad fields defines a class of reference frames
\cite{Maluf4}. If we denote by $x^\mu(s)$ the world line $C$ of an 
observer in space-time, and by $u^\mu(s)=dx^\mu/ds$ its velocity 
along $C$, we may identify the observer's velocity with the $a=(0)$ 
component of $e_a\,^\mu$ \cite{Hehl2}. Thus 
$u^\mu(s)=e_{(0)}\,^\mu$ along $C$.
The acceleration of the observer is given by 
$a^\mu= Du^\mu /ds =De_{(0)}\,^\mu /ds =
u^\alpha \nabla_\alpha e_{(0)}\,^\mu$, where the covariant derivative
is constructed out of the Christoffel symbols. We see that $e_a\,^\mu$
determines the velocity and acceleration along the worldline of an 
observer adapted to the frame. From this perspective we conclude that 
a given set of tetrad fields, for which $e_{(0)}\,^\mu$ describes a 
congruence of timelike curves, is adapted to a particular class of 
observers, namely, to observers characterized by the velocity field 
$u^\mu=e_{(0)}\,^\mu$, endowed with acceleration $a^\mu$. If 
$e^a\,_\mu \rightarrow \delta^a_\mu$ in the limit 
$r \rightarrow \infty$, then $e^a\,_\mu$ is adapted to static
observers at spacelike infinity. 

Actual calculations of the energy-momentum and angular momentum of 
the gravitational field in the configuration space of the theory 
require the evaluation of the right hand side of eqs. (9) and (15).
The definitions of $P^a$ and $L^{ab}$ above are well defined if we
consider tetrad fields $e^a\,_\mu$ such that in the flat space-time 
limit (i.e., in the absence of the gravitational field) we have 
$T_{a\mu\nu}(e)=0$. However there are flat space-time tetrads 
$E^a\,_\mu$ for which $T_{a\mu\nu}(E)\ne 0$. Consequently for such
tetrads we obtain nonvanishing values of $P^a$ and $L^{ab}$ 
in the absence of the gravitational field. Therefore these
expressions must be regularized. The regularization of the 
gravitational energy-momentum is discussed in detail in ref. 
\cite{Maluf4}. Conceptually it is the same regularization procedure
that takes place in the Brown-York method \cite{BY}, which may be 
understood as the subtraction of the flat space-time energy. We 
denote $T^a\,_{\mu\nu}(E)=
\partial_\mu E^a\,_\nu-\partial_\nu E^a\,_\mu$,
and $\Pi^{aj}(E)$ as the expression of $\Pi^{aj}$ constructed out of 
{\it flat tetrads} $E^a\,_\mu$. The regularized form of the 
gravitational energy-momentum $P^a$ is defined by

\begin{equation}
P^a=-\int_V d^3x\,\partial_k\lbrack\Pi^{ak}(e) - \Pi^{ak}(E)\rbrack\;.
\label{19}
\end{equation}
This definition guarantees that the energy-momentum of the flat 
space-time always vanishes. The reference space-time is determined by 
the tetrad fields $E^a\,_\mu$, obtained from $e^a\,_\mu$ by requiring
the vanishing of the physical parameters like mass, angular momentum, 
etc. The definition above for $P^a$ has been investigated in Ref. 
\cite{Maluf4} by considering a set of tetrad fields for the Kerr 
black hole for which $T_{a\mu\nu}(E)\ne 0$. By subtracting the flat
space-time quantity $\Pi^{ak}(E)$ we arrive at the expected value for 
the total gravitational energy of the Kerr space-time.
We remark that the regularization of the gravitational
energy-momentum has been addressed recently in the  
analysis of ref. \cite{OR}.

We may likewise establish the regularized expression for the
gravitational 4-angular momentum. It reads

\begin{equation}
L^{ab}=\int_V d^3x\; \lbrack M^{ab}(e) - M^{ab}(E) \rbrack\,.
\label{20}
\end{equation}
Expressions (19) and (20) allow the evaluation of the gravitational 
energy-momentum and 4-angular momentum out of an {\it arbitrary} set 
of tetrad fields. 

\section{The space-time of the rotating mass shel}

We will present in detail the application of
definition (20) to
the space-time of a slowly rotating sphericall mass shell,
formulated by Cohen \cite{Cohen}. It describes a 
mathematically simple, non-singular configuration of the 
gravitational field that exhibits rotational effects and is
everywhere regular. In the limit of small angular momentum the
metric for such space-time corresponds to the asymptotic form 
of Kerr's metric tensor. The main motivation \cite{Cohen} for
considering this metric is the construction of a realistic
source for the exterior region of the Kerr space-time, and
therefore to match the latter region to a singularity-free
space-time. For a shell of radius $r_0$ and total mass
$m=2\alpha$ as seen by an observer at infinity, the metric
reads

\begin{equation}
ds^2=-V^2dt^2+\psi^4\lbrack dr^2+r^2d\theta^2+
r^2\sin^2\theta(d\phi-\Omega dt)^2\rbrack\;,
\label{21}
\end{equation}
where

\begin{eqnarray}
V&=&{{ r_0-\alpha}\over{r_0 + \alpha}}\,, \nonumber \\
\psi&=& \psi_0 =1+ {\alpha \over r_0}\,, \nonumber \\
\Omega&=&\Omega_0={\em constant}\,,
\label{22}
\end{eqnarray}
for $r < r_0$, and

\begin{eqnarray}
V&=&{{r-\alpha}\over{r+\alpha}}\,, \nonumber \\
\psi&=&1+{\alpha \over r}\,, \nonumber \\
\Omega&=&\biggl({{r_0 \psi_0^2}\over{r \psi^2}}\biggr)^3
\Omega_0 \,,
\label{23}
\end{eqnarray}
for $r> r_0$. 

The metric tensor given by Eq. (21) is a solution of
Einstein's equations up yo first order in $\Omega$. We recall that
$\Omega_0$ is the dragging angular velocity of locally inertial 
observers inside the shell. The contravariant components of the 
metric tensor will be useful in the following considerations. 
They read

\begin{equation}
g^{\mu\nu}= \pmatrix{
-{1\over V^2}&0&0&-{\Omega \over V^2} \cr
0&{1\over \psi^4}&0&0\cr
0&0&{1\over {r^2\psi^4}}&0\cr
-{\Omega \over V^2}&0&0&{{V^2-r^2\Omega^2\psi^4 \sin^2\theta}\over
{V^2r^2\psi^4\sin^2\theta}}\cr}\,.
\label{24}
\end{equation}

We will consider two simple configurations of tetrad fields and
discuss their physical interpretation as reference frames. The first 
one is given by

\begin{equation}
e_{a\mu}=\pmatrix{ -V&0&0&0\cr
\Omega r\psi^2 \sin\theta\,\sin\phi &
\psi^2 \,\sin\theta\,\cos\phi &
r\psi^2\,\cos\theta\,\cos\phi &
-r\psi^2\,\sin\theta\,\sin\phi\cr
-\Omega r\psi^2 \sin\theta\,\cos\phi &
\psi^2 \,\sin\theta\,\sin\phi &
r\psi^2\,\cos\theta\,\sin\phi &
r\psi^2\,\sin\theta\,\cos\phi\cr
0 & \psi^2\,\cos\theta & -r\psi^2\,\sin\theta & 0 \cr}\,.
\label{25}
\end{equation}
The determinant of $e_{a\mu}$ is
$e=V r^2\psi^6\,\sin\theta$. We find that the tetrad fields above are
adapted to observers whose four-velocity in space-time is given by

\begin{equation}
e_{(0)}\,^\mu(t,r,\theta,\phi)=
{1\over V}(1,0,0,\Omega)\,.
\label{26}
\end{equation}
Therefore an observer at the radial position $r$ rotates around the 
source along a circular trajectory, with angular velocity 
$\Omega(r)$ ($1/V$ is the normalization factor of the 
four-velocity).

In order to obtain the spatial components of the gravitational 
angular momentum we need the evaluation of 
$T_{\lambda \mu\nu}= e^a\,_\lambda T_{a\mu\nu}$. We find

\begin{eqnarray}
T_{001}&=&V\partial_1V -{1\over 2}\partial_1(\Omega r\psi^2)^2
\sin^2\theta\,, \nonumber \\
T_{301}&=& r\psi^2 \partial_1(\Omega r\psi^2) \sin^2\theta\,,
\nonumber \\
T_{002}&=&-(\Omega r\psi^2)^2 \sin\theta\,\cos\theta\,,
\nonumber \\
T_{302}&=& \Omega r^2 \psi^4 \sin\theta\,\cos\theta\,,
\nonumber \\
T_{103}&=&-\Omega r \psi^4 \sin^2\theta\,, \nonumber \\
T_{203}&=& -\Omega r^2 \psi^4 \sin\theta\,\cos\theta\,, 
\nonumber \\
T_{212}&=& r^2 \psi^2 (\partial_1 \psi^2)\,,\nonumber \\
T_{013}&=& -\Omega r^2\psi^2 (\partial_1 \psi^2) \sin^2\theta\,,
\nonumber \\
T_{313}&=& r^2 \psi^2 (\partial_1 \psi^2) \sin\theta\,.
\label{27}
\end{eqnarray}
After long but simple calculations we obtain

\begin{eqnarray}
\Sigma_{001}&=& {1\over 2} (T_{001}-g_{00}T_1)\,, \nonumber \\
\Sigma_{301}&=& {1\over 4} (T_{301}-T_{013}+T_{103})
-{1\over 2} g_{03}T_1\,, \nonumber \\
\Sigma_{002}&=& {1\over 2} T_{002}\,, \nonumber \\
\Sigma_{103}&=& {1\over 4} (T_{103}+T_{013}+T_{301})\,,
\nonumber \\
\Sigma_{212}&=& {1\over 2} (T_{212} +g_{22}T_1)\,, \nonumber \\
\Sigma_{013}&=& {1\over 4} (T_{013}+T_{103}-T_{301})+
{1\over 2} g_{03}T_1\,, \nonumber \\
\Sigma_{313}&=& {1\over 2} (T_{313}+g_{33}T_1)\,, \nonumber \\
\Sigma_{023}&=& {1\over 2} T_{203}\,.
\label{28}
\end{eqnarray}
In view of the relations 
$e^{(1)}\,_\mu g^{\mu 0}=e^{(2)}\,_\mu g^{\mu 0}=0$, the expression
for $M^{(1)(2)}$ is simplified to

\begin{eqnarray}
M^{(1)(2)}&=&-4ke\,g^{11}\biggl[ 
(e^{(1)}\,_\mu g^{\mu 3})e^{(2)}\,_1 
-(e^{(2)}\,_\mu g^{\mu 3})e^{(1)}\,_1 \biggr]\times \nonumber \\
&{}& \times
\biggr[g^{00}\Sigma_{301}-g^{00} \Sigma_{103} -g^{03}\Sigma_{313}
\biggr]
\label{29}
\end{eqnarray}
It turns out that 
$g^{00}\Sigma_{301}-g^{00} \Sigma_{103} -g^{03}\Sigma_{313}=0$, and
therefore

\begin{equation}
M^{(1)(2)}=0\,.
\label{30}
\end{equation}
The other components also vanish, $M^{(1)(3)}=M^{(2)(3)}=0$, and
consequently the angular momentum $L^{(i)(j)}$
of the space-time of a rotating mass shell vanishes if computed out 
of Eq. (25). However this result should come as no surprise, since 
the reference frame represented by Eq. (25) is rotating around the 
source with angular velocity $\Omega$.

Now we address  another configuration of tetrad fields that 
has a simple interpretation as reference frame. We consider

\begin{equation}
e_{a\mu}=\pmatrix{-X&0&0&Z\cr
0&\psi^2\sin\theta\,\cos\phi&r\psi^2 \cos\theta\,\cos\phi&
-Y\sin\theta\,\sin\phi \cr
0&\psi^2\sin\theta\,\sin\phi&r\psi^2 \cos\theta\,\sin\phi&
Y\sin\theta\,\cos\phi \cr
0&\psi^2 \cos\theta &-r\psi^2 \sin\theta&0}\,,
\label{31}
\end{equation}
where

\begin{eqnarray}
X&=&(V^2-r^2 \Omega^2 \psi^4 \sin^2\theta)^{1/2}\,, \nonumber \\
Y&=& -{1\over X}\, \Omega r^2 \psi^4 \sin^2\theta \,, \nonumber \\
Z&=& {V\over X}\, r\psi^2\,.
\label{32}
\end{eqnarray}
This set of tetrad fields yields the velocity field given by

\begin{equation}
e_{(0)}\,^\mu(t,r,\theta,\phi)=
\biggl({1\over X},0,0,0 \biggr)\,.
\label{33}
\end{equation}
According to the physical interpretation of Eq. (31), the latter
is adapted to static observers in space-time.

The nonvanishing components of $T_{\lambda \mu\nu}$ are

\begin{eqnarray}
T_{001}&=& X\partial_1 X \,, \nonumber \\
T_{301}&=& -Z \partial_1 X \,, \nonumber \\
T_{202}&=& X\partial_2 X \,, \nonumber \\
T_{302}&=& -Z\partial_2 X \,, \nonumber \\
T_{212}&=& r^2\psi^2(\partial_1 \psi^2)\,, \nonumber \\
T_{013}&=& X\partial_1 Z \,, \nonumber \\
T_{313}&=& -Z\partial_1 Z+ (\partial_1 Y -
\psi^2)Y \sin^2\theta\,, \nonumber \\
T_{023}&=& X \partial_2 Z\,, \nonumber \\
T_{323}&=& -Z \partial_2 Z +Y(\partial_2 Y)\sin^2\theta-
(r\psi^2-Y)Y \sin\theta\,\cos\theta\,.
\label{34}
\end{eqnarray}
The quantites above yield the following nonvanishing components
of $\Sigma_{\lambda \mu\nu}$:

\begin{eqnarray}
\Sigma_{001}&=& {1\over 2}(T_{001}-g_{00}T_1)\,, \nonumber \\
\Sigma_{301}&=& {1\over 4}(T_{301}-T_{013})-
                {1\over 2} g_{03}T_1 \,, \nonumber \\
\Sigma_{002}&=& {1\over 2}(T_{002}-g_{00}T_2)\,, \nonumber \\
\Sigma_{302}&=& {1\over 4}(T_{302}-T_{023})-
                {1\over 2} g_{03}T_2\,, \nonumber \\
\Sigma_{103}&=& {1\over 4}(T_{013}+T_{301})\,, \nonumber \\
\Sigma_{112}&=& -{1\over 2}g_{11}T_2\,, \nonumber \\
\Sigma_{212}&=& {1\over 2}(T_{212}+g_{22}T_1)\,, \nonumber \\
\Sigma_{013}&=& {1\over 4}(T_{013}-T_{301})+{1\over 2}g_{03}T_1
\,, \nonumber \\
\Sigma_{313}&=& {1\over 2}(T_{313}+g_{33}T_1)\,, \nonumber \\
\Sigma_{023}&=& {1\over 4}(T_{023}-T_{302})+{1\over 2}
g_{03}T_2\,, \nonumber \\
\Sigma_{323}&=& {1\over 2}(T_{323}+g_{33}T_2)\,.
\label{35}
\end{eqnarray}
The traces of the torsion tensor are

$$ T_1=g^{00}T_{001} +g^{03}(T_{301}-T_{013})-g^{22}T_{212}
-g^{33}T_{313}\,,$$

$$T_2=g^{00}T_{002}+g^{03}(T_{302}-T_{023})-g^{33}T_{323}\,.$$

The only physical parameter in Eqs. (21) and (31) is $\alpha$, 
which is related to the mass of the shell by means of $m=2\alpha$.
By making $\alpha=0$ we find that $T_{(0)13}=\partial_1 Z$ and
$T_{(0)23}=\partial_2 Z$ are nonvanishing, and therefore
$T_{013}\ne 0$ and $T_{123}\ne 0$ in this limit. The latter
quantities behave as $O(r^{-2}\sin^2\theta)$ 
and $O(r^{-1}\sin\theta\,\cos\theta)$, respectively. 
Consequently we antecipate that it will be necessary
to make use of the regularized definition of the gravitational
algular momentum. 

The exact expression of $M^{(1)(2)}$ is given by

\begin{eqnarray}
M^{(1)(2)}&=&-2ke
(e^{(1)}\,_3 e^{(2)}\,_1-e^{(1)}\,_1 e^{(2)}\,_3)\times\nonumber \\
&{}& \times \biggr[
g^{00}g^{03}g^{11}(T_{001}-g_{00}T_1) \nonumber \\
&{}& -g^{00}g^{11}g^{33}(T_{013}+g_{03}T_1) \nonumber \\
&{}&+g^{03}g^{03}g^{11}(T_{301}-g_{03}T_1) \nonumber \\
&{}& -g^{03}g^{11}g^{33}(T_{313}+g_{33}T_1)\biggr] \nonumber \\
&{}& -2ke
(e^{(1)}\,_3e^{(2)}\,_2 -e^{(1)}\,_1e^{(2)}\,_3) \times \nonumber \\
&{}& \times \biggl[
g^{00}g^{03}g^{22}(T_{002}-g_{00}T_2) \nonumber \\
&{}& g^{00}g^{22}g^{33}\biggl(
{1\over 2} (T_{302}-T_{023})-g_{03}T_2\biggr) \nonumber \\
&{}& - g^{03}g^{03}g^{22}\biggl(
{1\over 2}(T_{023}-T_{302})+g_{03}T_2\biggr) \nonumber \\
&{}&-g^{03}g^{22}g^{33}(T_{323}+g_{33}T_2)\biggr]
\label{36}
\end{eqnarray}

In order to simplify the expression above we will make two
assumptions. We will assume

\begin{eqnarray}
r^2 \Omega^2 &<< & 1 \,, \\
r_0 & >> &  \alpha \,.
\label{37,38}
\end{eqnarray}
Equation (37) is necessary because the metric tensor given by
Eq. (21) is a solution of Einstein's equations in the limit of 
slow rotation. The assumption given 
by Eq. (38) simplifies several calculations, but we note that
it holds in the consideration of ordinary laboratory objects, for
which we know the value of the Newtonian angular momentum. Both
conditions imply that 
$X=(V^2-r^2 \Omega^2 \psi^4 \sin^2\theta)^{1/2}$ is always real.

Condition (37) simplifies Eq. (36) to

\begin{eqnarray}
M^{(1)(2)} &\cong & -2ke(e^{(1)}\,_3 e^{(2)}\,_1 -
e^{(1)}\,_1 e^{(2)}\,_3) \times \nonumber \\
&\times &\biggl[ g^{00}g^{03}g^{11}(T_{001}-g_{00}T_1)-
g^{00}g^{11}g^{33}T_{013} \biggr] \nonumber \\
&+& ke (e^{(1)}\,_3 e^{(2)}\,_2 - e^{(1)}\,_2 e^{(2)}\,_3)
g^{00}g^{22}g^{33}T_{023}\,.
\label{39}
\end{eqnarray}

Taking into account now condition (38) we obtain

\begin{equation}
M^{(1)(2)} \cong  -4k\biggl[
2\alpha \Omega r \sin^3\theta -{1\over 2} \Omega r^2 \sin^3\theta
+{1\over 2} \Omega r^2 \sin\theta\,\cos^2\theta\ \biggr]\,.
\label{40}
\end{equation}
Integration of the last two terms in the equation above diverges.
These terms arise precisely because of the $T_{013}$ and
$T_{023}$ components, and we see that they do not depend on the
parameter $\alpha$. Thus these terms must be regularized,
according to the discussion after Eq. (35). 

The first term in Eq. (40) arises due to the spatial derivative
of $\psi$ (specifically, there arises the term  
$\partial_1\psi^2=-(2\alpha)/r^2$
in the combination $T_{001}-g_{00}T_1$). Therefore the first
term vanishes for $r<r_0$ because $\partial_1 \psi^2=0$ in this
region, and we finally have

\begin{equation}
M^{(1)(2)}(\alpha) - M^{(1)(2)}(\alpha=0) \cong 
-{1\over {4\pi}} 2\alpha (\Omega r) \sin^3\theta \,,
\label{41}
\end{equation}
for $r>r_0$. In terms of the notation of Eq. (20) we have
$M^{(1)(2)}(\alpha) - M^{(1)(2)}(\alpha=0)=
M^{(1)(2)}(e) - M^{(1)(2)}(E)$. 

Integration of Eq. (41) yields

\begin{equation}
L^{(1)(2)} \cong -{{8\alpha} \over {3r_0}} J=
-{{4m}\over {3r_0}} J\,,
\label{42}
\end{equation}
where

\begin{equation}
J={1\over 2} (r_o\psi_0^2)^3 \Omega_0\,.
\label{43}
\end{equation}
It is not difficult to verify that

\begin{equation}
L^{(1)(3)}=L^{(2)(3)}=0\,.
\label{44}
\end{equation}

We arrive at precisely the same value obtained in Ref. 
\cite{Maluf1}, except for the sign, in the analysis of the same
gravitational field configuration. We recall 
that Cohen \cite{Cohen} identifies $J$ given by Eq. (43) as the
Newtonian value for the angular momentum of a rotating mass shell.
In similarity to the analysis of Ref. \cite{Maluf1}, it is possible
to write $L^{(1)(2)}$ as the product of the moment of 
inertia of the source with $\Omega_0$, which is the induced 
angular velocity of inertial frames inside the shell. More
precisely, we have \cite{Maluf1}

\begin{equation}
L^{(1)(2)}= -\biggl( {2\over 3} mr_0^2 \biggr) \Omega_0\,.
\label{45}
\end{equation}
We can take into account the whole discussion presented in
Ref. \cite{Maluf1} and assert that $L^{(1)(2)}$ yields the value
of the angular momentum of the field, not of the source (we have
to reintroduce the constants $c$ and $G$ in all definitions). 
For weak gravitational fields we expect the gravitational angular 
momentum to be of small intensity in laboratory (cgs) units.
The gravitational field of a mass shell of
typical laboratory values is negligible, and consequently we would
expect its gravitational angular momentum to be negligible as well.
In contrast, the gravitational angular momentum of the space-time of
a rotating mass shell obtained by means of the Komar integral
\cite{Komar} yields a value that is of the same order of magnitude 
of the angular momentum of the source \cite{Maluf1}. 

\section{Final remarks}

In this article we have addressed the gravitational angular momentum
in the framework of the TEGR. The definitions of energy-momentum and
4-angular momentum arise in the realm of the field equations of the 
TEGR and are separately invariant under general coordinate 
transformations of the three-dimensional space and under time
reparametrizations (according to the discussion at the end of section
2). The gravitational angular momentum is frame dependent, and in 
the two cases studied above such dependence is clearly consistent 
with the physical interpretation of the frames. The general 
characterization
of tetrad fields as reference frames in terms of the velocity and
acceleration fields, $u^\mu=e_{(0)}\,^\mu$ and $a^\mu$, respectively,
is an issue that must be better understood in general.

The gravitational 4-angular momentum in the context of the TEGR has 
already been investigated in Refs. \cite{BV,Nester}. These analyses 
are essentially of the same nature as those developed in Refs.
\cite{RG,York,BM,Sza}, namely, the gravitational angular momentum is
identified with Hamiltonian boundary terms at spacelike infinity.
In particular, the boundary terms considered in Refs. \cite{BV}
and \cite{Nester} agree with each other in the limit
$r \rightarrow \infty$.

It is very likely that the asymptotic boundary conditions  
and the parity conditions discussed by Beig and
\'{o} Murchadha \cite{BM} and by Szabados \cite{Sza} are necessary 
in order to arrive at a
finite value for the gravitational 4-angular momentum, even in the
context of the regularized definition given by Eq. (20). This issue
must be investigated. It is not clear that the regularization
procedure eliminates all types of infinities when the integrations 
are performed over the spacelike section of the space-time. 

The evaluation of the gravitational angular momentum of the 
space-time of a rotating black hole, given by the Kerr solution, is 
not an easy procedure. The attempt towards this problem, briefly
exposed in Ref. \cite{Maluf1}, led us to conclude that the essential 
singularity of the solution poses a problem in the integration of 
the angular momentum density. However, we believe that this problem 
may be overcome by adopting a more convenient set of coordinates 
\cite{Doran} to the Kerr solution in the context of the regularized
definition (20).

In view of Eq. (17) we define the gravitational Pauli-Lubanski 
vector $W_a$,

\begin{equation}
W_a={1\over 2} \varepsilon_{abcd} P^bL^{cd}\,.
\label{46}
\end{equation}
It is straightforward to verify that $W_a$ commutes with $P^b$,

\begin{equation}
\lbrace W_a , P^b \rbrace =0\,.
\label{47}
\end{equation}
Therefore we may also define the Casimir type quantities $P^2$ and
$W^2$,

\begin{eqnarray}
P^2&=& \eta_{ab}P^aP^b \,,\nonumber \\
W^2&=& \eta^{ab}W_aW_b\,,
\label{48}
\end{eqnarray}
which commute with both $P^a$ and $W_a$. In this sense we may say 
that $P^2$ and $W^2$ are invariants. These quantities might
play an important role in the characterization of the 
gravitational field. In ref. \cite{Maluf5} it was evaluated the
value of $P^2$ for a Schwarzschild black hole of mass $m$. It was 
obtained $P^2=-m^2c^2$ in the framework of a static or moving 
observer. The generality of this result must be investigated.
Another issue to be addressed is the polarization of
gravitational waves. It should be analyzed to what extent 
$W_a$ and $P^a$ yield information about the helicity of plane
gravitational waves, for instance. We recall that investigations
about the possible existence of spin-1 gravitational waves have 
been carried out in the literature \cite{CPV}. This issue will
considered in the context of the present analysis.

\end{document}